\documentclass[aps,prl,twocolumn,showpacs,preprintnumbers,amsmath,amssymb]{revtex4-1}
 \def\ep{{\epsilon}}

 \def\frac#1#2{{#1\over #2}}

 \def\s{\sqrt}

\def\be{\begin{equation}}
\def\ee{\end{equation}}
\def\ba{\begin{eqnarray}}
\def\ea{\end{eqnarray}}

 \def\de{\partial}

 \def\f {\frac}
 \def\ti{\tilde}

 \def\no{\nonumber \\}

 \def\ep{\epsilon}

 \def\vp{\varphi}

\usepackage{color}
\usepackage{graphicx}% Include figure files
\usepackage{dcolumn}% Align table columns on decimal point
\usepackage{bm}% bold math

\begin{document}

\title{Anti--de Sitter Space from Optimization of Path Integrals in Conformal Field Theories}
YITP-17-20 ; IPMU17-0039
\author{Pawel Caputa$^{a}$, Nilay Kundu$^{a}$, Masamichi Miyaji$^{a}$, 
Tadashi Takayanagi$^{a,b}$ and Kento Watanabe$^{a}$}

\affiliation{$^a$Center for Gravitational Physics, Yukawa Institute for Theoretical Physics,
Kyoto University, Kitashirakawa Oiwakecho, Sakyo-ku, Kyoto 606-8502, Japan}

\affiliation{$^{b}$Kavli Institute for the Physics and Mathematics of the Universe, University of Tokyo, Kashiwa, Chiba 277-8582, Japan}

\date{\today}

\begin{abstract}
We introduce a new optimization procedure for Euclidean path integrals which compute wave functionals in conformal field theories (CFTs). We optimize the background metric in the space on which the path integration is performed. Equivalently this is interpreted as a position-dependent UV cutoff. For two-dimensional CFT vacua, we find the optimized metric is given by that of a hyperbolic space and we interpret this as a continuous limit of the conjectured relation between tensor networks and Anti--de Sitter (AdS)/conformal field theory (CFT) correspondence. We confirm our procedure for excited states, the thermofield double state, the Sachdev-Ye-Kitaev model and discuss its extension to higher-dimensional CFTs. We also show that when applied to reduced density matrices, it reproduces entanglement wedges and holographic entanglement entropy. We suggest that our optimization prescription is analogous to the estimation of computational complexity.
\end{abstract}

%\pacs{72.10.-d,73.21.-b,73.50.Fq}
\maketitle

%\begin{figure}
%  \centering
%  \includegraphics[width=5cm]{SS.eps}
%  \caption{A sketch of Surface/State Correspondence in AdS/CFT.}
%\label{fig:ss}
%  \end{figure}

Recently, candidates on the basic mechanism of Anti--de Sitter (AdS)/conformal field theory (CFT) correspondence \cite{Ma} have been investigated actively. One prominent candidate is based on emergent spaces via tensor networks, such as MERA (multi-scale entanglement renormalization ansatz) \cite{MERA} as pioneered in \cite{Swingle}; for recent developments refer to e.g. \cite{cMERA,Beny,NRT,MT,HAPPY,Cz,MNSTW,HQ,MTW}. The holographic computation of entanglement entropy \cite{RT,HRT} is naturally explained by this idea.

The purpose of this Letter is to reformulate this conjectured connection to tensor networks from the viewpoint of Euclidean path integrals. Indeed, the tensor network renormalization \cite{TNR} shows an Euclidean path integral description of a vacuum wave functional is well approximated by a MERA tensor network. In this approach, we first discretize the path integral into a lattice and rewrite it as a tensor network. Next the network is optimized by contracting the tensors and lattice sites are reduced, finally leading to the MERA network.

In our approach we would like to keep working in the Euclidean path integral description and
perform the optimization by changing the structure (or geometry) of lattice regularization (refer to Fig.\ref{pathfig}), which was first considered in \cite{MTW}, as explained in appendix  \ref{appA}. In this Letter, we will present a systematic formulation of optimization by describing the change of regularization as that of the metric on the space where the path integral is done.
\begin{figure}
  \centering
  \includegraphics[width=6cm]{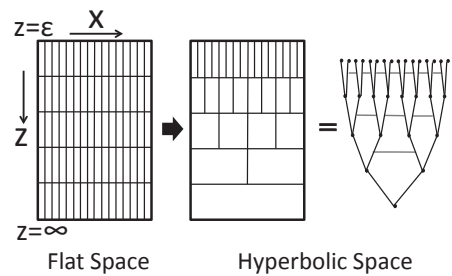}
  \caption{A computation of ground state wave functional from Euclidean path integral and its optimization, which is described by a hyperbolic geometry. The right figure schematically shows its tensor network expression.}
\label{pathfig}
  \end{figure}

 Our path integral approach has a number of advantages. Clearly, we can analyze any CFTs, including genuine holographic ones, while tensor network approaches provide toy models of holography.
 Also in tensor network descriptions there is a subtle issue that the MERA can also be interpreted as a de Sitter space \cite{Beny,Cz}, while the refined tensor networks in \cite{HAPPY,HQ} are argued to describe Euclidean hyperbolic spaces. In our Euclidean approach we can avoid this issue and we can directly show that the emergent space is a hyperbolic space.

The ground state wave functional in $d$-dimensional CFTs on $R^d$ is computed by
an Euclidean path integral:
\ba
&& \Psi_{\mbox{CFT}}(\ti{\vp}(x))=\int \left(\prod_{x}\prod_{\ep<z<\infty}D\vp(z,x)\right)e^{-S_{CFT}(\vp)}  \no
&& \hspace{2cm}\ \ \ \  \cdot \prod_{x}\delta(\vp(\ep,x)-\ti{\vp}(x)), \label{wfgr}
\ea
where the Euclidean time $\tau$ is related to $z$ via $z=-(\tau-\ep)$ and $x$ describes the other $d-1$-dimensional directions. We introduce a UV cutoff $\ep$ (i.e. the lattice constant) and consider a discretization of the above Euclidean path integral. The original flat space metric is given by
\be
ds^2=\ep^{-2}\cdot(dz^2+dx^idx^i),   \label{flat}
\ee
and one cell of the regularized lattice has the unit area.

Now we would like to optimize this Euclidean path integral by changing the geometry of lattice regularization as in Fig.\ref{pathfig}. The basic rule is to require that we should reproduce the correct vacuum wave functional (i.e. the one (\ref{wfgr}) for the metric (\ref{flat}) ) even after the optimization up to a normalization factor i.e. $\Psi_{\mbox{opt}}\propto\Psi_{\mbox{CFT}}$. The optimization can be done by modifying the background metric for the path integration as
\ba
&& ds^2=g_{zz}(z,x)dz^2+g_{ij}(z,x)dx^i dx^j+2g_{zj}(z,x)dzdx^j,  \no
&& g_{zz}(z=\ep,x)=\ep^{-2}, \ \ g_{ij}(z=\ep,x)=\delta_{ij}\cdot \ep^{-2}, \no
&& g_{iz}(z=\ep,x)=0,
\ea
where the constraints argue that the UV regularization agrees with the original one (\ref{flat}) at $z=\ep$ to reproduce the correct wave functional after the optimization.

Intuitively, our optimization corresponds to minimizing the number of lattice points, which is expected to measure the number of tensors (or complexity) in a tensor network. Note that we have in mind the procedure of tensor network renormalization \cite{TNR}, where a discretized path integral
is mapped into a tensor network which consists of unitaries and isometries (refer also to the full paper \cite{long}). It is not immediately clear what is the right measure for this in field theories. However, in the two-dimensional CFTs, we can explicitly identify this measure, owing to its simple structure, as we will see below.

Before we go on, we would like to point out that the above optimization procedure can be generalized to excited states by inserting operators in the middle of path integration. Also, we can equally optimize nonconformal field theories if we assume a UV fixed point. For example, we can deform a CFT by adding a (time-independent) external field $J(x)$. Then the optimization can be done by allowing a $z$ dependence of the source $J(z,x)$.

In two-dimensional CFTs, any metric is written as the diagonal form via a coordinate transformation 
\ba
&& ds^2=e^{2\phi(z,x)}(dz^2+dx^2), \ \ e^{2\phi(z=\ep,x)}=\frac{1}{\ep^2}. \label{met}
\ea
The function $\phi(z,x)$ describes the metric. With the universal UV cutoff $\ep$, the measure of quantum fields $\vp$ in the CFT anomalously changes under Weyl rescaling \cite{GM}
\ba
[D\vp]_{g_{ab}=e^{2\phi}\delta_{ab}}=e^{S_L[\phi]-S_L[0]}\cdot [D\vp]_{g_{ab}=\delta_{ab}},
\ea
where $S_L[\phi]$ is the Liouville action \cite{Po} (see also \cite{GM,Wadia}),
\be
S_L[\phi]\!=\!\frac{c}{24\pi}\!\int^\infty_{-\infty}\!\! dx\! \int^\infty_{\ep}\!\! dz\!\! \left[ (\de_x \phi)^2\!+\!(\de_z \phi)^2\!+\!\mu e^{2\phi}\!+\!R_0\phi\right], \label{lac}
\ee
where $R_0$ is the Ricci scalar of the original space, which is zero in our setup here.
The constant $c$ is the central charge of two-dimensional CFT. The kinetic term in $S_L$ describes the conformal anomaly and the potential term comes from UV regularization. In our treatment, $\mu$ is an $O(1)$ constant and we will simply set $\mu=1$ by an appropriate shift of $\phi$. For earlier relations between Louville theory and three-dimensional gravity, refer to,  e.g., \cite{lth}, whose connections to our Letter are not obvious.

The ground-state wave functional $\Psi_{g_{ab}=e^{2\phi}\delta_{ab}}$ computed from the path integral for the metric (\ref{met}) is proportional to the one $\Psi_{g_{ab}=\delta_{ab}}$ for the flat metric owing the conformal symmetry with the coefficient
\ba
\Psi_{g_{ab}=e^{2\phi}\delta_{ab}}(\ti{\vp}(x))=e^{S_L[\phi]-S_L[0]}\cdot \Psi_{g_{ab}=\delta_{ab}}(\ti{\vp}(x)).
\ea

Now we argue that the optimization is equivalent to minimizing the normalization $e^{S_L[\phi]}$ of the wave functional. The reason is that this factor measures the number of repetitions of the same operation (i.e. the path integral over a cell). In other words, the optimization chooses the most efficient Euclidean path integral.

Thus, the optimization can be completed by requiring the equation of motion of Liouville action $S_L$
\be
4\de_w \de_{\bar{w}}\phi=e^{2\phi},   \label{leom}
\ee
where we introduced $w=z+ix$ and $\bar{w}=z-ix$.
Its general solution is given by (see, e.g., \cite{Se,GM})
\be
e^{2\phi}=\frac{4A'(w)B'(\bar{w})}{(1-A(w)B(\bar{w}))^2}.  \label{abf}
\ee
Note that functions $A(w)$ and $B(\bar{w})$ describe the degrees of freedom of conformal mappings.

First we choose the solution: $A(w)=w,\ \  B(\bar{w})=-1/\bar{w}.$
This leads to the Poincare metric $H_2$ 
\be
e^{2\phi}=4(w+\bar{w})^{-2}=z^{-2}.  \label{hyp}
\ee
We argue that this is interpreted as the time slice of AdS$_3$ dual to a holographic CFT. This is consistent with the tensor network description of AdS/CFT and can be regarded as its continuous version. Notice that we did not fix the overall normalization of the metric or equally the AdS radius $R_{AdS}$ because in our formulation, it depends on the precise definition of UV cutoff. However, we can apply the argument of \cite{MTW} and can heuristically argue that $R_{AdS}$ is proportional to the central charge $c$.

This solution is the minimum of $S_L$ with the boundary condition
$e^{2\phi(z=\ep,x)}=\ep^{-2}$. For this, we rewrite (\ref{lac}) into
\ba
&& S_L=\frac{c}{24\pi}\int dxdz \left[(\de_x\phi)^2+(\de_z\phi+e^{\phi})^2\right]
\no
&& -\frac{c}{12\pi}\int dx [e^{\phi}]^{z=\infty}_{z=\ep}\geq \frac{cL}{12\pi\ep},
\label{newe}
\ea
where $L\equiv \int dx$ is the size of space direction and we assumed the IR behavior
$e^{2\phi(z=\infty,x)}=0$. The final inequality in (\ref{newe}) is saturated iff
$\de_x\phi=\de_z\phi+e^{\phi}=0$, and this indeed leads to the solution (\ref{hyp}).

It is also interesting to note that $S_L$ scales like $S_L\sim c\frac{L}{\ep}$, and this
agrees with the behavior of the computational complexity \cite{Susskind} of a CFT ground-state and the quantum information metric \cite{InfoM} for the same state,
both of which are given by the volume of the time slice of AdS. Indeed, our minimization of $S_L$ matches with the optimization of the quantum circuits, which defines the complexity.

Let us turn to the setup of a CFT on a cylinder, where the wave functional is defined on a circle $|w|=1$ at a fixed Euclidean time. After the optimization procedure, we obtain the geometry $e^{2\phi(w,\bar{w})}=4(1-|w|^{2})^{-2}$, which is precisely the Poincare disk and is the solution to (\ref{leom}).

 Now we consider an excited state given by a primary state. The excitation is described by a primary operator $O(w,\bar{w})$ with the conformal dimension $h=\bar{h}$. Its behavior under the Weyl rescaling is expressed as $O(w,\bar{w})\propto e^{-2h \phi}.$
When we insert the operator $O$ at the origin, the dependence of the wave function on $\phi$ reads
\ba
\Psi_{g_{ab}=e^{2\phi}\delta_{ab}}\simeq e^{S_L[\phi]-S[0]}  e^{-2h\phi(0)} \Psi_{g_{ab}=\delta_{ab}}.
\ea

Consider a CFT$_2$ on a disk $|w|<1$ and insert a source at the center $w=0$. The equation of motion
becomes
\be
4\de_w \de_{\bar{w}}\phi-e^{2\phi}+2\pi(1-a)\delta^2(w)=0
\label{eomf},
\ee
where we set
\be
a=1-12h/c.  \label{dila}
\ee
We can focus on the solutions $A(w)=w^a,\ B(\bar{w})=\bar{w}^a.$
The metric looks like
\be
e^{2\phi}=\frac{4a^2}{|w|^{2(1-a)}(1-|w|^{2a})^2}. \label{defcit}
\ee
Since the angle of $w$ coordinate is $2\pi$ periodic, this geometry has the deficit angle $2\pi(1-a)$.

Let us compare this with the time slice of the gravity dual predicted from AdS$_3/$CFT$_2$. Indeed it is given by the deficit angle geometry (\ref{defcit}) with the identification
\be
a=\s{1-24h/c}.  \label{three}
\ee
Thus, the geometry from our optimization (\ref{dila}) agrees with the gravity dual (\ref{three}) when $h\ll c$, where the back reaction of the point particle is very small. This argument can also be generalized into an excitation at a generic point on the disk simply by acting the $SL(2,R)$ symmetry of the AdS$_3$ which preserves the time slice.

However, if we assume the quantum Liouville theory rather than the classical one, we find a perfect matching. In the quantum Liouville theory, we introduce a parameter $\gamma$ such that $c=1+3Q^2$ with $Q\equiv\frac{2}{\gamma}+\gamma$. The chiral conformal dimension of the operator $e^{\frac{2\beta}{\gamma}\phi}$ is $\frac{\beta(Q-\beta)}{2}$. If $c$ is large enough to have a classical gravity dual, we get
$a\simeq 1-\beta\gamma\simeq \s{1-\frac{24h}{c}},$
 which agrees with (\ref{three}).

This agreement suggests that the actual optimized wave functional is given by a ``quantum" optimization \ba
\Psi_{\mbox{opt}}[\ti{\vp}]=\left[\int D\phi(z,x)e^{-S_L[\phi]}\biggr{|}_{\phi_{z=\ep}=\ti{\vp}}\!\!\!\cdot\!\!\left(\Psi_{g_{ab}=\delta_{ab}}[\ti{\vp}]\right)^{-1}\right]^{-1}.
\label{qwe}
\ea
If we take the semiclassical approximation, we reproduce our classical optimization by minimizing
$S_L$.

We can extend our analysis to a finite temperature $T=1/\beta$ case. In the thermofield double description \cite{MaE}, the wave functional is computed from a path integral on a cylinder with a finite width $-{\beta \over 4}(\equiv z_{1})< z< {\beta \over 4}(\equiv z_2)$ in the Euclidean time direction, more explicitly
\ba
&& \Psi[\ti{\vp}_1(x),\ti{\vp}_2(x)]=\int \left(\prod_{x}\prod_{-{\beta \over 4}<z<{\beta \over 4}}D\vp(z,x)\right)e^{-S_{CFT}(\vp)} \no
&& \times\!\!\prod_{-\infty<x<\infty}\!\!\!\delta\big(\vp\left(z_1,x\right)\!-\!\ti{\vp}_1(x)\big)
~\delta\big(\vp\left(z_2,x\right)\!-\!\ti{\vp}_2(x)\big),
\ea
where $\ti{\vp}_1(x)$ and $\ti{\vp}_2(x)$ are the boundary values for the fields of the CFT (i.e. $\ti{\vp}(x)$) at $z= \mp {\beta \over 4}$, respectively.

By minimizing the Liouville action $S_L$, we reach the solution in (\ref{abf}) given by:
$A(w)=e^{{2 \pi i w\over \beta}},\ \  B(\bar{w})=-e^{{2 \pi i \bar{w} \over \beta}}.$
This leads to
\be \label{phforbtz}
e^{2\ti\phi}=\frac{16 \pi^2}{\beta^2}\frac{e^{{2 \pi i \over \beta}(w+\bar{w})}}{\left(1+ e^{{2 \pi i \over \beta}(w+\bar{w})}\right)^2}=\frac{4 \pi^2}{\beta^2} \sec^2 \left({2 \pi z\over \beta}\right).
\ee
This precisely agrees with the time slice of eternal Banados-Teitelboim-Zanelli (BTZ) black hole  (i.e. the Einstein-Rosen
bridge) \cite{MaE}.

\begin{figure}
  \centering
  \includegraphics[width=6cm]{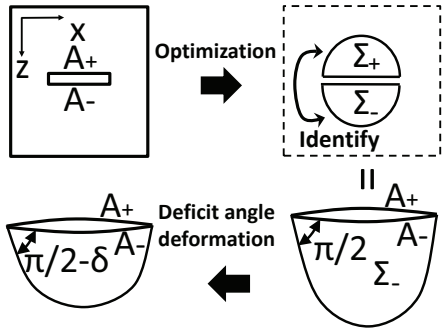}
  \caption{The optimization of path integral for a reduced density matrix. The upper left picture is the definition of $\rho_A$ in terms of the path integral in flat space. The upper right one is the one after the optimization and is equivalent to a geometry which is obtained by pasting two identical entanglement wedges along the geodesic (=the half circle) as shown in the lower right picture. If we start from $\rho_A^n$, we obtain the geometry in the lower left picture with $\delta=\pi (1-n)$.}
\label{redfig}
  \end{figure}

Consider an optimization of path integral representation of reduced density matrix $\rho_A$. The subsystem $A$ is chosen as an interval $-l\leq x\leq  l$. $\rho_A$ is defined from the CFT vacuum by tracing out the complement of $A$ (the upper left picture in Fig.\ref{redfig}). The optimization is done by introducing the background metric as in (\ref{met}) where the Dirichlet boundary condition $e^\phi=1/\ep$ is imposed around the upper and lower edges of the slit $A$. The optimization squeezes the infinitely extended plane into a finite-size region, corresponding to contracting tensors in tensor networks. Finally the geometry is obtained by pasting the two regions $\Sigma_{\pm}$ along the boundaries $\de\Sigma_{\pm}$ (the upper right picture in Fig.\ref{redfig}). The extremization of the bulk action $S_L$ leads to the hyperbolic metric (\ref{hyp}).

The shape of the boundaries $\de\Sigma_{\pm}$ (we excluded the edges along $A$) is fixed by extremizing the boundary action in the Liouville theory \cite{FZZ}
\be
S_{Lb}=\frac{c}{12\pi}\int_{\de\Sigma_{\pm}} ds [K_0\phi+\mu_B e^{\phi}],  \label{bLv}
\ee
where $K_0$ is the (trace of) extrinsic curvature of the boundary $\de\Sigma_{\pm}$ in the flat space. The final term is the boundary Liouville potential. Since $\Sigma_+$ and $\Sigma_{-}$ are pasted along the boundary smoothly, we set $\mu_B=0$ (actually $\mu_B$ is related to the tension parameter in the gravity dual of a CFT on a manifold with boundaries \cite{Ta}).
By imposing the Neumann boundary condition of $\phi$, we find the extrinsic curvature $K$ in the curved metric (\ref{met}) is vanishing $e^{\phi}K=(n^x\de_x+n^z\de_z)\phi+K_0=0$,
where $n^{x,z}$ is the unit vector normal to the boundary in the flat space (see appendix (B) for details). Thus we find $K=0$ i.e. the condition of a geodesic and the boundary $\de \Sigma_{+}=\de \Sigma_{-}$  is given by the half circle $x^2+z^2=l^2$ in the hyperbolic space (\ref{hyp}). Interestingly this geometry obtained from the optimization of $\rho_A$, coincides with the entanglement wedge \cite{RT,HRT,EW1,EW2,EW3}, given by $x^2+z^2\leq l^2$.

Let us evaluate the entanglement entropy. Consider an optimization of the matrix product $\rho_A^n$.
We assume an analytical continuation of $n$ with $|n-1|\ll 1$.
The standard replica method leads to a conical deficit angle $2\pi(1-n)$
at the two end points of the interval $A$. Thus, after the optimization, we get a geometry with the corner angle $\pi/2+\pi(n-1)$ instead of $\pi/2$ (the lower left picture in Fig. \ref{redfig}). This modification of the boundary $\de\Sigma_{\pm}$ is equivalent to shifting the extrinsic curvature from $K=0$ to $K=\pi(n-1)$ (see appendix B). Since the boundary condition of $\phi$ is given by
$K+\mu_B=0$, this shift is realized by turning on $\mu_B$ as $\mu_B=\pi(1-n)$.

In the presence of infinitesimally small $\mu_B$ we can evaluate the Liouville action by a probe approximation neglecting all back reactions. By taking a derivative with respect to $n$, we obtain the entanglement entropy $S_A$
\ba
S_A&=&-\de_{n}\left[\frac{c\mu_B}{12\pi}\int_{\de\Sigma_{+}}ds\ e^{\phi} +\frac{c\mu_B}{12\pi}\int_{\de\Sigma_{-}}ds\ e^{\phi}\right]_{n=1}\no
&=& \frac{c}{6}\int_{\de\Sigma_{+}}ds\ e^{\phi}=\frac{c}{3}\log\frac{2 l}{\ep},\label{entd}
\ea
reproducing the well-known result \cite{HLW}. The lower left expression (\ref{entd}) $\frac{c}{6}\int_{\de\Sigma_{+}} e^\phi$
precisely agrees with the holographic entanglement entropy formula \cite{RT,HRT} as $\de\Sigma_{+}(=\de\Sigma_{-})$ has to be the geodesic. Refer also to the appendix (C) for another derivation of entanglement entropy.

As recently discovered, to make sense of AdS$_2/$CFT$_1$, we need a conformal symmetry breaking \cite{Kit,SY,SYK,Dilaton}, written by a Schwartzian derivative action as explicitly realized in
the Sachdev-Ye-Kitaev (SYK) model. For the one-dimensional metric $ds^2=e^{2\phi}d\tau^2$, the Schwartzian derivative term reads $N\int d\tau (\de_\tau\phi)^2$, where $N$ is a constant proportional to degrees of freedom. Thus, we find
\ba
&& \Psi_{g_{\tau\tau}=e^{2\phi}}(\ti{\vp}(x))=e^{S_{1}[\phi]-S_1[0]}\cdot \Psi_{g_{\tau\tau}=1}(\ti{\vp}(x)),  \no
&& S_{1}[\phi]=N\int d\tau\left[(\de_\tau\phi)^2+\mu e^{\phi}\right].
\ea
The action minimization leads to
$ds^2=e^{2\phi}d\tau^2=\frac{d\tau^2}{\tau^2}.$
This is consistent with the time slice of AdS$_2$ spacetime. Note if there were no conformal symmetry breaking effect, we could not stabilize the optimization procedure.

Higher-dimensional generalizations of our optimization procedure are not
straightforward as the generic metric cannot be expressed only by
the Weyl scaling like (\ref{met}).  Nevertheless, let us consider
what optimization can lead to correct time slices of gravity duals by taking into account only the Weyl scaling degrees of freedom. We argue for the metric (\ref{met}) with $x$ regarded as $d-1$-dimensional vector that the optimization can be done by minimizing the functional
(again $N$ describes degrees of freedom)
\be
S_{d}=N\int dx^{d-1}dz \left[e^{d\phi}+e^{(d-2)\phi}\left((\de_x\phi)^2+(\de_z\phi)^2\right)\right].\label{acd}
\ee
Indeed, the minimization of $S_d$ leads to the hyperbolic space H$_d$ which is the time slice of pure AdS$_{d+1}$. Moreover, one parameter deformation of H$_d$ is also a solution to the equation of motion for $\phi$ and this deformed metric matches with that of the time slice of the AdS Schwarzschild black hole up to the first-order perturbation (details are described in the full paper \cite{long}). For higher-order deformations, we expect quantum effects as in the AdS$_3/$CFT$_2$ case mentioned previously as in (\ref{qwe}). Also, for the H$_d$ solution, we can estimate $S_d\sim NV_{d-1}/\ep^{d-1}$, and this again agrees with the amount of complexity \cite{Susskind} and information metric \cite{InfoM}.

In this Letter, we proposed an optimization procedure of the Euclidean path integral for
quantum states in CFTs and gave an explicit formulation in terms of background metrics for two-dimensional CFTs. The optimization leads to the geometry which coincides with the time slice of its gravity dual. We argue that this provides a continuous version of the tensor network interpretation of AdS/CFT. When applied to a reduced density matrix, we naturally reproduce the entanglement wedge and the holographic entanglement entropy. At the same time, this gives a framework which calculates the computational complexity of quantum states in CFTs. Also this correspondence can be applied to the SYK model.

Moreover, we made a proposal on its higher-dimensional generalization and performed minimal checks.
Another interesting future problem is to develop our formalism to calculate correlation functions in CFTs. This will also be important to extend our method to nonconformal field theories. It is also important to extend our formalism to
time-dependent backgrounds. These detailed studies are currently ongoing and we hope to report them in future publications.\\

We thank Rajesh Gopakumar, Kanato Goto, Yasuaki Hikida, Veronika Hubeny, Alex Maloney, Yu Nakayama, Tatsuma Nishioka, Yasunori Nomura, Tokiro Numasawa, Hirosi Ooguri, Fernando Pastawski, Mukund Rangamani, Shinsei Ryu, Andy Strominger, Brian Swingle, Tomonori Ugajin, Herman Verlinde, Spenta Wadia, and Kazuya Yonekura for useful conversations.  M.M. and K.W. are supported by JSPS fellowships. P.C. and T.T. are supported by the Simons Foundation through the ``It from Qubit'' collaboration. N.K. and T.T. are supported by JSPS Grant-in-Aid for Scientific Research (A) No.16H02182. TT is also supported by World Premier International Research Center Initiative (WPI Initiative) from the Japan Ministry of Education, Culture, Sports, Science and Technology (MEXT). P.C., M.M., T.T. and K.W. thank very much the long term workshop ``Quantum Information in String Theory and Many-body Systems'' held at YITP, Kyoto where this work was initiated. T.T. is grateful to the conference ``Recent Developments in Fields, Strings, and Gravity'' held in Quantum Mathematics and Physics (QMAP) at UC Davis, where this work was presented. T.T. also thank the international symposium ``Frontiers in Mathematical Physics'' in Rikkyo U. and the meeting ``String Theory: Past and Present (Spenta Fest)'' in ICTS, Bangalore, where the present work was announced.

\section{Appendix A: Relation to Path Integral Representations of Tensor Networks} \label{appA}
In this appendix, we explain how our optimization procedure looks like in the path integral representation of tensor networks considered in \cite{MTW}. The Euclidean action is deformed keeping the ground state wave function fixed by introducing the position dependent UV cut off.
 Using this deformed action, we obtain a flow of wave functions by terminating the path-integral
 at various finite values of $z$ (see Fig.1), from the ground state to
an unentangled state. This flow is very similar to the continuous MERA tensor network
\cite{cMERA,NRT}. As in tensor networks, this deformation is far from unique.
 Our optimization procedure eliminates this ambiguity by minimizing action in the bulk.
  For simplicity, we take a two dimensional free real scalar field $f$ as an illustration.
As in \cite{MTW}, the on-shell action with UV cutoff $1/\epsilon$ is given by
 \be S[f]=\f{1}{2}\int^{\infty}_{\epsilon}dz\int \f{dk}{2\pi}~\Gamma(\epsilon|k|)\cdot |k|^2e^{-2|k|z}f(k)f(-k),
 \ee
 where $\Gamma(x)$ is a cut off function such that $\Gamma(x)=0~(x>1)$ and $\Gamma(x)=1~(x<1)$; $f(k)$ is the Fourier transform of $f$ at $z=\epsilon$.
 After $x$ independent Weyl rescaling $ds^2=dz^2+dx^2\rightarrow e^{2\phi(z)}(dz^2+dx^2)$, the on-shell action becomes
  \be S[f,\phi]=\f{1}{2}\int^{\infty}_{\epsilon}dz\int \f{dk}{2\pi}~ \Gamma(e^{-\phi(z)}|k|)\cdot|k|^2e^{-2|k|z}f(k)f(-k).
   \ee
  From the minimization of the Liouville action, we obtain $e^{2\phi}=\f{1}{\mu}\cdot z^{-2}$. As a result, we find
   the deformed on-shell action with coordinate dependent cut off
   \be \Gamma(e^{-\phi(z)}|k|)=\Gamma(z\cdot \sqrt{\mu}\cdot|k|),  \ee
which agrees with the result based on the heuristic argument in \cite{MTW}.

\section{Appendix B: Extrinsic Curvatures} \label{appB}

Consider a boundary $x=f(z)$ in the two dimensional space defined by the metric $ds^2=e^{2\phi(z,x)}(dz^2+dx^2)$. The out-going normal unit vector $N^a$ is given by
\ba
&& N^z=e^{-\phi(z,x)}n^z=\frac{-f'(z)e^{-\phi(z,x)}}{\s{1+f'(z)^2}}, \no
&& N^x=e^{-\phi(z,x)}n^x=\frac{e^{-\phi(z,x)}}{\s{1+f'(z)^2}},
\ea
where $n^a$ is the normal unit vector in the flat space $ds^2=dz^2+dx^2$.
The extrinsic curvature (=its trace part) at the boundary is defined by $K=h^{ab}\nabla_a N_b$, where all components are projected to the boundary whose induced metric is written as $h_{ab}$.
Explicitly we can calculate $K$ as follows:
\ba
K&=&\frac{e^{-\phi(z,x)}}{\s{1+f'(z)^2}}\left[\de_x\phi-f'\de_z\phi-\f{f''}{1+(f')^2}\right]\no
&=&e^{-\phi(z,x)}\left[n^a\de_a\phi+K_0\right],
\ea
where $K_0=-\frac{f''}{(1+(f')^2)^{3/2}}$ is the extrinsic curvature of the boundary $x=f(z)$ in the flat metric $ds^2=dz^2+dx^2$. Note that in the hyperbolic space $\phi=-\log z+$const., the circle $z^2+x^2=l^2$ is a solution to $K=0$ for any values of $l$.

Note that by setting the variation of the action with respect to the infinitesimal shift of $\phi$ at the boundary for the total Liouville action $S_L+S_{Lb}$ in the presence of $\mu_B$ of (\ref{bLv}), we get the boundary condition \be
K+\mu_B=e^{-\phi(z,x)}\left[n^a\de_a\phi+K_0\right]+\mu_B=0.  \label{ncb}
\ee
Therefore the turning on $\mu_B$ means the boundary condition with the
non-zero extrinsic curvature. Note that along the cut on the subsystem $A$ at $z=\ep$, we impose the Dirichlet boundary condition $e^\phi=1/\ep$. On the other boundary of $\de\Sigma_{\pm}$ (refer to Fig.\ref{redfig}, i.e. on the half circle), we impose the Neumann boundary condition (\ref{ncb}).

If we consider the boundary given by $x^2+(z-z_0)^2=l^2$, we get $K=z_0/l$.
When $z_0$ is infinitesimally small, we get $x\simeq l+(z_0/l)\cdot z+O(z^2)$ near the boundary point
$(z,x)=(0,l)$. Therefore the corner angle is found to be $\pi/2-\delta$ with $\delta\simeq -z_0/l$
(for the definition of $\delta$, refer also to lower pictures in Fig.\ref{redfig}).
Therefore we find the relation $K\simeq -\delta$. For the $n$-sheeted replica
geometry used in the main context of the letter, we chose $\mu_B=-K\simeq \delta=\pi(1-n)$.\\

\section{Appendix C: Another Derivation of Entanglement Entropy} \label{appC}

Here we would like to present another computation of entanglement entropy.
As in the standard replica method, this leads to the conical deficit angle $2\pi(1-n)$
at the two end points of the interval $A$. Thus we have the contribution from the scalar curvature term in (\ref{lac}): $\int_{\Sigma_{\pm}} R_0=4\pi(1-n)$. Since we know that the topology of the path-integrated space for $\rho_A^n$ is the same as $\rho_A$ i.e. the disk, we find $\int_{\de\Sigma_{\pm}}K_0=2\pi(1-n)$ so that the total Euler number does not depend on $n$:
$\frac{1}{4\pi}\sum_{\eta=\pm}\left[\int_{\Sigma_{\eta}} R_0+2\int_{\Sigma_{\eta}}K_0\right]=0$.
By evaluating the full Liouville action $S_L+S_{Lb}$, the terms which are proportional to $(n-1)$ are found as
\ba
\!\!\!\!\frac{c}{24\pi}\!\sum_{\eta=\pm}\!\left[\!\int_{\Sigma_\eta}\!R_0\phi \! +\!2\int_{\de\Sigma_{\eta}}\!K_0\phi\!\right]
\!\simeq\! \frac{c}{3}\!(1\!-\!n)\!\log\frac{l}{\ep}.
\ea
Therefore the entanglement entropy is evaluated as
\be
S_A=-\frac{\de (S_L+S_{Lb})}{\de n}|_{n=1}\simeq \frac{c}{3}\log\frac{l}{\ep},
\ee
reproducing the well-known result \cite{HLW}.

%It is also intriguing to note that the same result
%is obtained by introducing infinitesimally small $\mu_B$ such that $\mu_B=2\pi(1-n)$ on the boundary %$\de\Sigma_{+}$. Indeed its contribution to $S_A$ is proportional to the geodesic length:
%$S_A=\frac{c}{6}\int_{\de\Sigma_{+}} e^\phi$, which agrees with the holographic entanglement entropy %\cite{RT,HRT}.


\begin{thebibliography}{99}


\bibitem{Ma}
  J.~M.~Maldacena,
  Adv.\ Theor.\ Math.\ Phys.\  {\bf 2} (1998) 231
  [Int.\ J.\ Theor.\ Phys.\  {\bf 38} (1999) 1113]
  [arXiv:hep-th/9711200];
  %%CITATION = IJTPB,38,1113;%%



\bibitem{MERA}
G.~Vidal,
%``A class of quantum many-body states that can be efficiently simulated,''
Phys. Rev. Lett. {\bf 101}, 110501 (2008) , arXiv:quant-ph/0610099;
%``Entanglement renormalization,''
Phys. Rev. Lett. {\bf 99}, 220405 (2007) , arXiv:cond-mat/0512165.




\bibitem{Swingle}
  B.~Swingle,
%``Entanglement Renormalization and Holography,''
Phys. Rev. {\bf D 86}, 065007 (2012), arXiv:0905.1317 [cond-mat.str-el].  %%CITATION = ARXIV:0905.1317;%%

\bibitem{cMERA}
  J.~Haegeman, T.~J.~Osborne, H.~Verschelde and F.~Verstraete,
%``Entanglement Renormalization for Quantum Fields in Real Space,''
  Phys.\ Rev.\ Lett.\  {\bf 110}, no. 10, 100402 (2013)
  [arXiv:1102.5524 [hep-th]].
  %%CITATION = doi:10.1103/PhysRevLett.110.100402;%%

\bibitem{Beny}
  C.~Beny,
%``Causal structure of the entanglement renormalization ansatz,''
  New J.\ Phys.\  {\bf 15} (2013) 023020
  doi:10.1088/1367-2630/15/2/023020
  [arXiv:1110.4872 [quant-ph]].
  %%CITATION = doi:10.1088/1367-2630/15/2/023020;%%


\bibitem{NRT}
  M.~Nozaki, S.~Ryu and T.~Takayanagi,
%``Holographic Geometry of Entanglement Renormalization in Quantum Field Theories,''
JHEP {\bf 1210} (2012) 193  doi:10.1007/JHEP10(2012)193  [arXiv:1208.3469 [hep-th]].  %%CITATION = doi:10.1007/JHEP10(2012)193;%%


\bibitem{MT}
M.~Miyaji and T.~Takayanagi,
%``Surface/State Correspondence as a Generalized Holography,''
PTEP {\bf 2015} (2015) 7,  073B03  doi:10.1093/ptep/ptv089  [arXiv:1503.03542 [hep-th]].  %%CITATION = doi:10.1093/ptep/ptv089;%%


\bibitem{HAPPY}
 F.~Pastawski, B.~Yoshida, D.~Harlow and J.~Preskill,
 %``Holographic quantum error-correcting codes: Toy models for the bulk/boundary correspondence,''
  JHEP {\bf 1506} (2015) 149
  doi:10.1007/JHEP06(2015)149
  [arXiv:1503.06237 [hep-th]].
  %%CITATION = doi:10.1007/JHEP06(2015)149;%%



\bibitem{Cz}
 B.~Czech, L.~Lamprou, S.~McCandlish and J.~Sully,
%``Integral Geometry and Holography,''
  JHEP {\bf 1510} (2015) 175
  doi:10.1007/JHEP10(2015)175
  [arXiv:1505.05515 [hep-th]].
  %%CITATION = doi:10.1007/JHEP10(2015)175;%%

\bibitem{MNSTW}
  M.~Miyaji, T.~Numasawa, N.~Shiba, T.~Takayanagi and K.~Watanabe,
%``Continuous Multiscale Entanglement Renormalization Ansatz as Holographic Surface-State %Correspondence,''
Phys.\ Rev.\ Lett.\  {\bf 115} (2015) 17,  171602  doi:10.1103/PhysRevLett.115.171602  [arXiv:1506.01353 [hep-th]].  %%CITATION = doi:10.1103/PhysRevLett.115.171602;%%


\bibitem{HQ}
  P.~Hayden, S.~Nezami, X.~L.~Qi, N.~Thomas, M.~Walter and Z.~Yang,
%``Holographic duality from random tensor networks,''
  JHEP {\bf 1611}, 009 (2016)
  doi:10.1007/JHEP11(2016)009
  [arXiv:1601.01694 [hep-th]].
  %%CITATION = doi:10.1007/JHEP11(2016)009;%%


\bibitem{MTW}
  M.~Miyaji, T.~Takayanagi and K.~Watanabe,
%``From Path Integrals to Tensor Networks for AdS/CFT,''
 Phys.\ Rev.\ D {\bf 95} (2017) no.6,  066004, arXiv:1609.04645 [hep-th].

\bibitem{RT}
  S.~Ryu and T.~Takayanagi,
%``Holographic derivation of entanglement entropy from AdS/CFT,''
Phys.\ Rev.\ Lett.\  {\bf 96} (2006) 181602  [hep-th/0603001].  %%CITATION = doi:10.1103/PhysRevLett.96.181602;%%

\bibitem{HRT}
  V.~E.~Hubeny, M.~Rangamani and T.~Takayanagi,
%``A Covariant holographic entanglement entropy proposal,''
JHEP {\bf 0707} (2007) 062  [arXiv:0705.0016 [hep-th]].  %%CITATION = ARXIV:0705.0016;%%


\bibitem{TNR}
G.~Evenbly and G.~Vidal,
%`` Tensor Network Renormalization ,''
arXiv:1412.0732 [cond-mat.str-el],
 Phys. Rev. Lett. 115, 180405 (2015);
%`` Tensor network renormalization yields the multi-scale entanglement renormalization ansatz, ''
arXiv:1502.05385 [cond-mat.str-el],  Phys. Rev. Lett. 115, 200401 (2015).

\bibitem{long}
P.~Caputa, N.~Kundu, M.~Miyaji, T.~Takayanagi and K.~Watanabe,
  %``Liouville Action as path integral Complexity: From Continuous Tensor Networks to AdS/CFT,''
 arXiv:1706.07056 [hep-th].
 %%CITATION = ARXIV:1706.07056;%%
 
\bibitem{GM}
  P.~H.~Ginsparg and G.~W.~Moore,
%``Lectures on 2-D gravity and 2-D string theory,''
hep-th/9304011.  %%CITATION = HEP-TH/9304011;%%


\bibitem{Po}
  A.~M.~Polyakov,
%``Quantum Geometry of Bosonic Strings,''
Phys.\ Lett.\  {\bf 103B} (1981) 207.
%%CITATION = doi:10.1016/0370-2693(81)90743-7;%%

\bibitem{Wadia}
S.~R.~Das, S.~Naik and S.~R.~Wadia,
%``Quantization of the Liouville Mode and String Theory,''
Mod.\ Phys.\ Lett.\ A {\bf 4} (1989) 1033.
%%CITATION = doi:10.1142/S0217732389001209;%%
%135 citations counted in INSPIRE as of 19 Feb 2017

\bibitem{lth}
 O.~Coussaert, M.~Henneaux and P.~van Driel,
%``The Asymptotic dynamics of three-dimensional Einstein gravity with a negative cosmological %constant,''
Class.\ Quant.\ Grav.\  {\bf 12} (1995) 2961 [gr-qc/9506019];  %%CITATION = doi:10.1088/0264-9381/12/12/012;%%
K.~Krasnov,
%``3-D gravity, point particles and Liouville theory,''
Class.\ Quant.\ Grav.\  {\bf 18} (2001) 1291 [hep-th/0008253].  %%CITATION = doi:10.1088/0264-9381/18/7/311;%%

\bibitem{Se}
  N.~Seiberg,
%``Notes on quantum Liouville theory and quantum gravity,''
Prog.\ Theor.\ Phys.\ Suppl.\  {\bf 102} (1990) 319.  %%CITATION = doi:10.1143/PTPS.102.319;%%


\bibitem{Susskind}

L.~Susskind,
%``Computational Complexity and Black Hole Horizons,''
Fortsch.\ Phys.\  {\bf 64} (2016) 24 [arXiv:1403.5695 [hep-th];
  %%CITATION = doi:10.1002/prop.201500092;%%
D.~Stanford and L.~Susskind,
  %``Complexity and Shock Wave Geometries,''
Phys.\ Rev.\ D {\bf 90} (2014) no.12,  126007  [arXiv:1406.2678 [hep-th]];
 %%CITATION = doi:10.1103/PhysRevD.90.126007;%%
A.~R.~Brown, D.~A.~Roberts, L.~Susskind, B.~Swingle and Y.~Zhao,
%``Holographic Complexity Equals Bulk Action?,''
Phys.\ Rev.\ Lett.\  {\bf 116} (2016) no.19,  191301
[arXiv:1509.07876 [hep-th]].
  %%CITATION = doi:10.1103/PhysRevLett.116.191301;%%



\bibitem{InfoM}
M.~Miyaji, T.~Numasawa, N.~Shiba, T.~Takayanagi and K.~Watanabe,
%``Distance between Quantum States and Gauge-Gravity Duality,''
Phys.\ Rev.\ Lett.\  {\bf 115} (2015) no.26,  261602
[arXiv:1507.07555 [hep-th]].
%%CITATION = doi:10.1103/PhysRevLett.115.261602;%%


\bibitem{MaE}
J.~M.~Maldacena,
%``Eternal black holes in anti-de Sitter,''
JHEP {\bf 0304} (2003) 021  doi:10.1088/1126-6708/2003/04/021  [hep-th/0106112].
%%CITATION = doi:10.1088/1126-6708/2003/04/021;%%



\bibitem{FZZ}
V.~Fateev, A.~B.~Zamolodchikov and A.~B.~Zamolodchikov,
%``Boundary Liouville field theory. 1. Boundary state and boundary two point function,''
hep-th/0001012.  %%CITATION = HEP-TH/0001012;%%



\bibitem{Ta}
 T.~Takayanagi,
%``Holographic Dual of BCFT,''
Phys.\ Rev.\ Lett.\  {\bf 107} (2011) 101602 [arXiv:1105.5165 [hep-th]].  %%CITATION = doi:10.1103/PhysRevLett.107.101602;%%


\bibitem{EW1}
B.~Czech, J.~L.~Karczmarek, F.~Nogueira and M.~Van Raamsdonk,
%``The Gravity Dual of a Density Matrix,''
Class.\ Quant.\ Grav.\  {\bf 29} (2012) 155009 [arXiv:1204.1330 [hep-th]].  %%CITATION = doi:10.1088/0264-9381/29/15/155009;%%

\bibitem{EW2}
A.~C.~Wall,
%``Maximin Surfaces, and the Strong Subadditivity of the Covariant Holographic Entanglement Entropy,''
Class.\ Quant.\ Grav.\  {\bf 31} (2014) no.22,  225007  [arXiv:1211.3494 [hep-th]].  %%CITATION = doi:10.1088/0264-9381/31/22/225007;%%

\bibitem{EW3}
  M.~Headrick, V.~E.~Hubeny, A.~Lawrence and M.~Rangamani,
%``Causality and holographic entanglement entropy,''
JHEP {\bf 1412} (2014) 162  doi:10.1007/JHEP12(2014)162  [arXiv:1408.6300 [hep-th]].  %%CITATION = doi:10.1007/JHEP12(2014)162;%%

\bibitem{HLW}
C.~Holzhey, F.~Larsen and F.~Wilczek,
%``Geometric and renormalized entropy in conformal field theory,''
Nucl.\ Phys.\ B {\bf 424} (1994) 443 [hep-th/9403108].
  %%CITATION = doi:10.1016/0550-3213(94)90402-2;%%

\bibitem{Kit}
A. Kitaev, “A simple model of quantum holography.”
Talks at KITP, April 7, 2015 and May 27, 2015.

\bibitem{SY}
 S. Sachdev and J.-w. Ye,
 %“Gapless spin ﬂuid ground state in a random, quantum Heisenberg magnet,”
 Phys.\ Rev.\ Lett. {\bf 70} (1993) 3339, arXiv:cond-mat/9212030 [cond-mat].


\bibitem{SYK}
J.~Maldacena and D.~Stanford,
%``Remarks on the Sachdev-Ye-Kitaev model,''
Phys.\ Rev.\ D {\bf 94} (2016) no.10,  106002
[arXiv:1604.07818 [hep-th]].
%%CITATION = doi:10.1103/PhysRevD.94.106002;%%

\bibitem{Dilaton}
J.~Maldacena, D.~Stanford and Z.~Yang,
PTEP {\bf 2016}, no. 12, 12C104 (2016)
[arXiv:1606.01857 [hep-th]].
%%CITATION = doi:10.1093/ptep/ptw124;%%






%%%%%%%%%%%%%%%%%%%%%%%%%%%%%%%%%%%%%%%%










\end{thebibliography}
\end{document}